\documentclass[prd,twocolumn,showpacs,nofootinbib, aps,10pt,superscriptaddress]{revtex4-1}

\usepackage{graphicx}
\usepackage{dcolumn}
\usepackage{bm}
\usepackage{amsmath}
\usepackage{amssymb}
\usepackage{xcolor}
\usepackage{hyperref}
\usepackage[capitalise]{cleveref}
\usepackage[activate={true,nocompatibility},final,kerning=true,factor=1100,stretch=10,shrink=10]{microtype}

\newcommand{\lcdm}{$\Lambda$CDM}
\newcommand{\dd}{\mathrm{d}}
\newcommand{\e}[1]{\mathrm{e}^{#1}}
\newcommand{\flrw}{Friedmann--Lema\^itre--Robertson--Walker}
\newcommand{\CLASS}{{\tt CLASS}}
\crefname{chapter}{chap.}{chap.}
\crefname{section}{Sec.}{Sec.}
\Crefname{chapter}{Chapter}{Chapters}
\Crefname{section}{Section}{Sections}

\newcommand{\eV}{\text{e\kern-0.15ex V}}

\newcommand{\TeV}{\text{T\kern-0.1ex\eV}}

\let\oldsqrt\sqrt
\def\sqrt{\mathpalette\DHLhksqrt}
\def\DHLhksqrt#1#2{
\setbox0=\hbox{$#1\oldsqrt{#2\,}$}\dimen0=\ht0
\advance\dimen0-0.2\ht0
\setbox2=\hbox{\vrule height\ht0 depth -\dimen0}
{\box0\lower0.4pt\box2}}

\newcommand{\ana}{Astron. Astrophys.} 

\newcommand{\apjs}{Astrophys. J., Suppl. Ser.}

\newcommand{\epj}{Eur. Phys. J.}

\newcommand{\jcap}{JCAP}

\newcommand{\mnras}{Mon. Not. Roy. Astron. Soc.}

\newcommand{\prev}{Phys. Rev.}
 
\newcommand{\plett}{Phys. Lett.}

\newcommand{\rpp}{Rept. Prog. Phys.}

\usepackage{tikz}
\usetikzlibrary{shapes.geometric, arrows}
\tikzstyle{chainstep} = [rectangle, rounded corners, minimum width=3cm, minimum height=0.8cm,text centered, draw=black, fill=white]

\begin{document}

\title{Understanding the suppression of structure formation from dark matter--dark energy momentum coupling}

\author{Finlay Noble Chamings}
 \email{finlay.noblechamings@nottingham.ac.uk}
\affiliation{%
 School of Physics and Astronomy, University of Nottingham
 \\Nottingham NG7 2RD, United Kingdom
}%
\author{Anastasios Avgoustidis}%
\email{Anastasios.Avgoustidis@nottingham.ac.uk}
\affiliation{%
 School of Physics and Astronomy, University of Nottingham
 \\Nottingham NG7 2RD, United Kingdom
}%
\author{Edmund J. Copeland}
\email{Edmund.Copeland@nottingham.ac.uk}
\affiliation{%
 School of Physics and Astronomy, University of Nottingham
 \\Nottingham NG7 2RD, United Kingdom
}%
\author{Anne M. Green}
\email{anne.green@nottingham.ac.uk}
\affiliation{%
 School of Physics and Astronomy, University of Nottingham
 \\Nottingham NG7 2RD, United Kingdom
}%
\author{Alkistis Pourtsidou}
\email{a.pourtsidou@gmul.ac.uk}
\affiliation{
 School of Physics and Astronomy, Queen Mary University of London,
 \\London E1 4NS, United Kingdom
}%

\date{\today}\title{Understanding the suppression of structure formation from dark matter--dark energy momentum coupling}

\begin{abstract}
Models in which scalar field dark energy interacts with dark matter via a pure momentum coupling have previously been found to potentially ease the structure formation tension between early- and late-universe observations. In this article we explore the physical mechanism underlying this feature. We argue analytically that the perturbation growth equations imply the suppression of structure growth, illustrating our discussion with numerical calculations. 
Then we generalise the previously studied quadratic coupling between the dark energy and dark matter to a more general power law case, also allowing for the slope of the dark energy exponential potential to vary. 
We find that the structure growth suppression is a generic feature of power law couplings and it can, for a range of parameter values, be larger than previously found.  
\end{abstract}

\pacs{Valid PACS appear here}

\maketitle

\section{\label{sec:intro}Introduction}

During the past two decades, cosmological observations have achieved a remarkable degree of precision. Measurements of Type~Ia supernovae~\cite{Riess:2016jrr}, the cosmic microwave background (CMB)~\cite{Planck2018VI}, and large scale structure~\cite{SDSS14,Samushia:2013yga}, indicate that around $96\%$ of the energy content of the universe is in the form of so-called dark energy and dark matter. These exotic species may be described by the standard cosmological model, \lcdm{}, in which dark energy takes the form of a cosmological constant and dark matter is taken to be cold, in other words having an equation of state equal to zero.

While \lcdm{} fits the available data very well, it suffers from a number of issues that motivate the study of alternatives. These include the fine-tuning~\cite{Martin:2012bt} and coincidence~\cite{whynow} problems. In addition, there are certain tensions between early- and late-universe observations in \lcdm{}. The present-day expansion rate of the universe, $H_0$ and the growth of structure, quantified by $\sigma_8$, can be calculated using the best-fit \lcdm{} parameters to cosmological data, including the CMB.
This gives rise to a smaller $H_0$ and a larger $\sigma_8$ than the results of local, late-universe measurements (for a recent discussion see Ref.~\cite{Verde:2019ivm}).
At present the tension in $H_0$ appears to be the more problematic of the two, though either or both issues may be caused by systematic effects that have not been accounted for. Future data from surveys such as EUCLID should confirm or resolve these tensions~\cite{Amendola:2012ys}. In the meantime it is worth exploring alternative explanations involving new physics. 
In this work we are especially interested in possible resolutions to the $\sigma_8$ tension.
The value of $\sigma_8$ inferred from CMB data is $0.811\pm 0.006$~\cite{Planck2018VI}, while cluster counts from the SZ effect give $\sigma_8 = 0.77\pm 0.02$~\cite{Ade:2013lmv} and weak lensing gives values of $\sigma_8$ ranging from $0.65$ to $0.75$~\cite{Hildebrandt:2016iqg, 10.1093/mnras/stx1820, 10.1093/mnras/stw3161}. 

A popular class of modifications to \lcdm{} is quintessence~\cite{1988PhRvD..37.3406R}, in which the cosmological constant $\Lambda$ is set to zero and a scalar field $\phi$ is introduced whose dynamical properties produce a negative equation of state giving rise to the observed late-time accelerated expansion of the universe. Normally it is assumed that the scalar field does not interact with dark matter. However there is no reason why this must be the case, and the consequences of relaxing this assumption have been widely studied. See Ref.~\cite{Wang:2016lxa} and references therein for a discussion of recent research on interacting dark energy.

Traditionally, couplings between dark energy and dark matter are introduced at the level of the equations of motion, for example:
\begin{equation}
\label{eq:tradIDE}
    \nabla^\mu T_{\mu\nu}^{(\text{c})} = J_\nu \,, \quad \quad \nabla^\mu T_{\mu\nu}^{(\text{DE})} = -J_\nu \,,
\end{equation}
such that the overall energy--momentum tensor $T_{\mu\nu} =  T_{\mu\nu}^{(\text{c})} + T_{\mu\nu}^{(\text{DE})}$, where $({\rm c})$ denotes cold dark matter and $({\rm DE})$ dark energy, is conserved as usual. $J_\nu$ is the flow of energy and momentum between dark energy and dark matter. 
A notable example was pioneered by Wetterich and Amendola~\cite{Wetterich:1994bg, Amendola:1999qq, Amendola:1999er}, in which $J_\nu = \beta T^{(\text{c})} \nabla_\nu \phi$, where $\beta$ is a constant, $\phi$ is the quintessence field and $T^{(\text{c})}$ is the trace of the dark matter energy--momentum tensor. Other couplings that have been proposed in the literature include promoting $\beta$ to be a function of $\phi$~\cite{Liu:2005wga, 2010PhRvD..82l3525L}, introducing a direct dependence on the expansion rate~\cite{Billyard:2000bh, Boehmer:2008av}, and couplings with non-linear dependence on the energy--momentum tensor or the scalar field gradient~\cite{Mimoso:2005bv, Chen:2008pz}. 

In Ref.~\cite{Pourtsidou:2013nha}, a construction was developed using the pull-back formalism for fluids to introduce dark energy--dark matter couplings at the level of the action. Defining the coupling at the level of the action is desirable for several reasons. It is often a more intuitive way to see the coupling, and it is easier to connect it to more fundamental physics. Perhaps more importantly, instabilities can often be more easily identified and avoided, saving time and computation when studying new models. The construction of Ref.~\cite{Pourtsidou:2013nha} leads to three distinct classes, or `Types' of models. Type~1 has been shown to include the commonly considered coupled quintessence model~\cite{Amendola:1999qq,Amendola:1999er} as a sub-class~\cite{Pourtsidou:2013nha,Skordis:2015yra}. Type~2 models have not been widely studied, but allow for both energy and momentum transfer between dark energy and dark matter. We focus on Type~3 models, which have a pure momentum coupling between dark energy and dark matter.

Type~3 models are interesting for several reasons. Due to the absence of a coupling at the background level between dark energy and dark matter, they are much less tightly constrained than Types~1 and~2~\cite{Pourtsidou:2013nha}. They give rise to a varying speed of sound of dark energy, the consequences of which were studied in Ref.~\cite{Linton:2017ged}. Perhaps most significantly, Type~3 models have been shown to provide a basis for easing the tension between early- and late-universe probes of structure formation by reducing the predicted value of $\sigma_8$ inferred from early-universe data~\cite{Pourtsidou:2016ico}.

In this paper we investigate the mechanism by which Type~3 models provide this reduction in structure growth and also study a more general form of the coupling function than previously considered. In \cref{sec:eqs} we present the cosmological equations of motion for the Type~3 models under consideration. In \cref{sec:overview} we describe in broad terms the way in which the structure growth suppression comes about, before explaining in detail the impact of a Type~3 coupling on the background cosmological evolution in \cref{sec:back} and how the linear perturbations are affected in \cref{sec:perts}. Finally in \cref{sec:disc} we present our conclusions and discuss possible avenues for future work.


\section{Equations of motion}
\label{sec:eqs}

In the formalism of Ref.~\cite{Pourtsidou:2013nha}, a Type~3 model is described by the Lagrangian:
\begin{equation}
\label{eq:LT3}
    L(n,Y,Z,\phi) = F(Y,Z,\phi) + f(n)\,.
\end{equation}
where $n$ is the fluid number density, $Y=(1/2)\nabla_\mu \phi \nabla^\mu\phi$ is the usual kinetic term, and
\begin{equation}
    Z = u^\mu \nabla_\mu \phi \,,
\end{equation}
is a direct coupling between the gradient of the scalar field and the fluid velocity $u^\mu$.

We consider a coupled quintessence model of the form:
\begin{equation}
    F = Y + V(\phi) + \gamma(Z)\,,
\end{equation}
where $V(\phi)$ is the scalar field potential and $\gamma(Z)$ is the coupling function. In this work we limit our analysis to power-law couplings of the form $\gamma(Z) = \beta_{n-2}Z^n$, where $n$ is an integer and $n \geq 2$. The background equations of motion may be found by assuming a spatially flat \flrw{} metric:
\begin{equation}
    \dd s^2 = a^2(\tau)(-\dd \tau^2 + \dd x_i \dd x^i)\,,
\end{equation}
where $a(\tau)$ is the scale factor and $\tau$ is the conformal time. $a$ evolves according to the usual Friedmann equation:
\begin{equation}
    \mathcal{H}^2 = \frac{1}{3 M_\text{P}^2} (\bar{\rho}_\text{b} + \bar{\rho}_\text{c} + \bar{\rho}_\gamma + \bar{\rho}_\phi)a^2\,,
\end{equation}
expressing the conformal Hubble parameter $\mathcal{H}$ in terms of the background energy densities of baryons ($\bar{\rho}_\text{b}$), dark matter ($\bar{\rho}_\text{c}$), radiation ($\bar{\rho}_\gamma$) and the scalar field ($\bar{\rho}_\phi$). $M_\text{P}$ denotes the reduced Planck mass. The energy density and pressure of the scalar field are given by~\cite{Pourtsidou:2013nha}
\begin{equation}
\label{eq:rhophi}
    \bar{\rho}_\phi = \frac{1}{2}\frac{\dot{\bar{\phi}}^2}{a^2} + \frac{\dot{\bar{\phi}}}{a}\gamma_{,Z} + \gamma(Z) + V(\phi)\,,
\end{equation}
\begin{equation}
\label{eq:pphi}
    \bar{p}_\phi = \frac{1}{2}\frac{\dot{\bar{\phi}}^2}{a^2} - \gamma(Z) - V(\phi)\,,
\end{equation}
where $\bar{\phi}$ is the background value of the scalar field and dots denote differentiation with respect to conformal time. The background part of $Z$ is given by $\bar{Z} = -\dot{\bar{\phi}}/a$. The scalar field obeys 
\begin{equation}
\label{eq:sfeback}
    (1-\gamma_{,ZZ})(\ddot{\bar{\phi}} - \mathcal{H}\dot{\bar{\phi}}) + 3a\mathcal{H}(\gamma_{,Z} - \bar{Z}) + a^2 V_{,\phi} = 0\,,
\end{equation}
and the background energy density of the cold dark matter is not modified by the Type~3 coupling:
\begin{equation}
    \dot{\bar{\rho}}_\text{c} + 3\mathcal{H}\bar{\rho}_\text{c} = 0\,.
\end{equation}

To perturb the equations of motion to linear order we work in the synchronous gauge, where the metric tensor reads:
\begin{equation}
    \dd s^2 = a^2(\tau)\left\{-\dd \tau^2 + \left[\left(1+\frac{1}{3}h\right)\gamma_{ij} + D_{ij}\nu\right]\dd x^i \dd x^j\right\}\,.
\end{equation}
Here, $h$ and $\nu$ are scalar perturbation variables, $D_{ij}$ is the traceless derivative operator $D_{ij} = \vec{\nabla}_i \vec{\nabla}_j - (1/3)\vec{\nabla}^2\gamma_{ij}$ and $\vec{\nabla}_i$ is the covariant derivative associated with the 3-space metric $\gamma_{ij}$.
The unit time-like vector field $u_\mu$ is perturbed as
\begin{equation}
    u_\mu = a(1,\vec{\nabla}_i\theta)\,.
\end{equation}
The cold dark matter density contrast $\delta_\text{c} = \delta\rho_\text{c}/\bar{\rho}_\text{c}$ obeys the standard continuity equation (in Fourier space):
\begin{equation}
\label{eq:delta0}
    \dot{\delta}_\text{c} = -k^2 \theta_\text{c} - \frac{1}{2}\dot{h}\,,
\end{equation}
while the velocity divergence $\theta_\text{c}$ obeys the modified Euler equation~\cite{Pourtsidou:2013nha}:
\begin{equation}
\label{eq:theta0}
    \dot{\theta}_\text{c} + \mathcal{H}\theta_\text{c} = \frac{(3\mathcal{H}\gamma_{,Z} + \gamma_{,ZZ}\dot{\bar{Z}})\delta\phi + \gamma_{,Z}\dot{\delta\phi}}{a(\bar{\rho}_\text{c} - \bar{Z}\gamma_{,Z})}\,,
\end{equation}
and the scalar field perturbation, $\delta\phi$, obeys
\begin{multline}
\label{eq:sfeperturb}
    (1-\gamma_{,ZZ})(\ddot{\delta\phi} + 2\mathcal{H}\dot{\delta\phi}) - \gamma_{,ZZZ}\dot{\bar{Z}}\dot{\delta\phi} 
    \\+ (k^2 + a^2V_{,\phi\phi})\delta\phi + \frac{1}{2}(\dot{\bar{\phi}} + a\gamma_{,Z})\dot{h} + a k^2 \gamma_{,Z} \theta_\text{c} = 0\,.
\end{multline}

The perturbed Einstein field equations are not modified by a Type~3 coupling and take their standard form, see e.g. Ref.~\cite{Ma:1995ey}.


\section{Overview of suppression of structure formation}
\label{sec:overview}

As found in Ref.~\cite{Pourtsidou:2016ico}, Type~3 models can result in a suppression of structure growth relative to \lcdm{}. We examine the mechanism by which this suppression occurs with reference to the underlying equations of motion. Following Ref.~\cite{Pourtsidou:2016ico}, we consider a Type~3 coupled quintessence model with an exponential potential of the form
\begin{equation}
    V(\phi) = A \e{-\lambda\phi/M_\text{P}}\,,
\end{equation}
and a power-law Type~3 coupling function given by
\begin{equation}
    \gamma(Z) = \beta_{n-2}Z^n\,,
\end{equation}
where for $n=2$ we recover the quadratic coupling studied in Ref.~\cite{Pourtsidou:2016ico}. 
We consider only values of $\beta_{n-2}$ such that $\gamma(Z)$ is negative, as positive $\gamma(Z)$ can result in a ghost instability if $\gamma_{,ZZ}>1$~\cite{Pourtsidou:2013nha}. Since $Z$ always takes negative values, this means that we consider only negative values of $\beta_{n-2}$ for $n$ even and only positive $\beta_{n-2}$ for $n$ odd.
We use the modified version of the Boltzmann code \CLASS{}~\cite{2011arXiv1104.2932L, 2011JCAP...07..034B, 2011arXiv1104.2934L, 2011JCAP...09..032L} developed by the authors of Ref.~\cite{Pourtsidou:2016ico}, further modifying it to compute the evolution of power-law couplings with $n>2$.

The matter power spectrum at a time $t$ is given by
\begin{equation}
    \label{eq:P}
    P(k,t) = \frac{2\pi^2}{k^3} T^2(k,t) \mathcal{P}(k)\,,
\end{equation}
where $\mathcal{P}(k)$ is the primordial power spectrum, which is assumed to have the form $\mathcal{P}(k) = A_\text{s}(k/k_*)^{n_\text{s}-1}$, and $T(k,t)$ is the transfer function describing the evolution of the matter density perturbation $\delta_\text{m}(k,t)$~\cite{Eisenstein:1997ik}.
All perturbed quantities, and the transfer function, are computed numerically by \CLASS{}.
The present-day matter power spectrum $P(k,t_0)$ is denoted by $P(k)$ for compactness. Since the primordial power spectrum is close to scale-invariant, with $n_\text{s}\approx 1$~\cite{Planck2018VI}, the matter power spectrum $P(k)$ derives all its interesting features from the transfer function. 
Due to the gravitational interaction between dark matter and baryons, their density contrasts obey $\delta_\text{c}\approx \delta_\text{b}\approx \delta_\text{m}$ to a very good approximation. In our numerical evolution we took $n_\text{s} = 0.97$~\cite{Planck2018VI}. 

The amplitude of the late-time matter density perturbations is commonly parameterised in terms of $\sigma_8$, defined as 
\begin{equation}
\label{eq:sigma}
    \sigma_R^2 = \frac{1}{2\pi^2}\int W_R(k)^2 P(k) k^2\dd k\,,
\end{equation}
with $R=8 h^{-1}\mathrm{Mpc}$, where $W_R(k)$ is the Fourier transform of the spherical top-hat window function:
\begin{equation}
    W_R(k) = \frac{3}{k^3R^3} \,[\sin(kR) - kR \cos(kR)]\,.
\end{equation}

\begin{figure}
    \centering
    \includegraphics[width=\columnwidth]{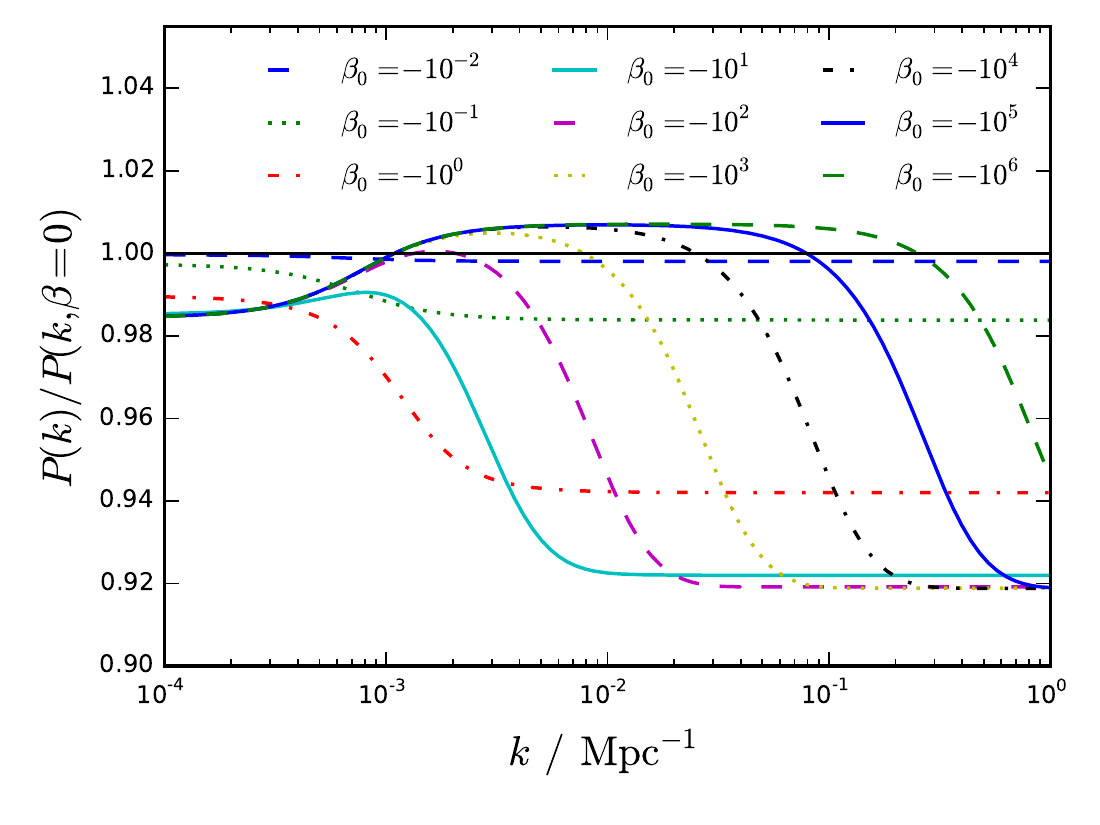}
    \caption{The linear matter power spectrum, $P(k)$,
    for an exponential potential, $V(\phi) = A \e{-\lambda\phi/M_\text{P}}$ and a coupling $\gamma(Z)=\beta_0 Z^2$ for various values of $\beta_{0}$.  The slope of the potential is set to $\lambda = 1.22$ and the sound horizon at recombination is held fixed at $\theta_\text{s} = 0.0104$.
    }
    \label{fig:0pk}
\end{figure}

\begin{figure}
    \centering
    \includegraphics[width=\columnwidth]{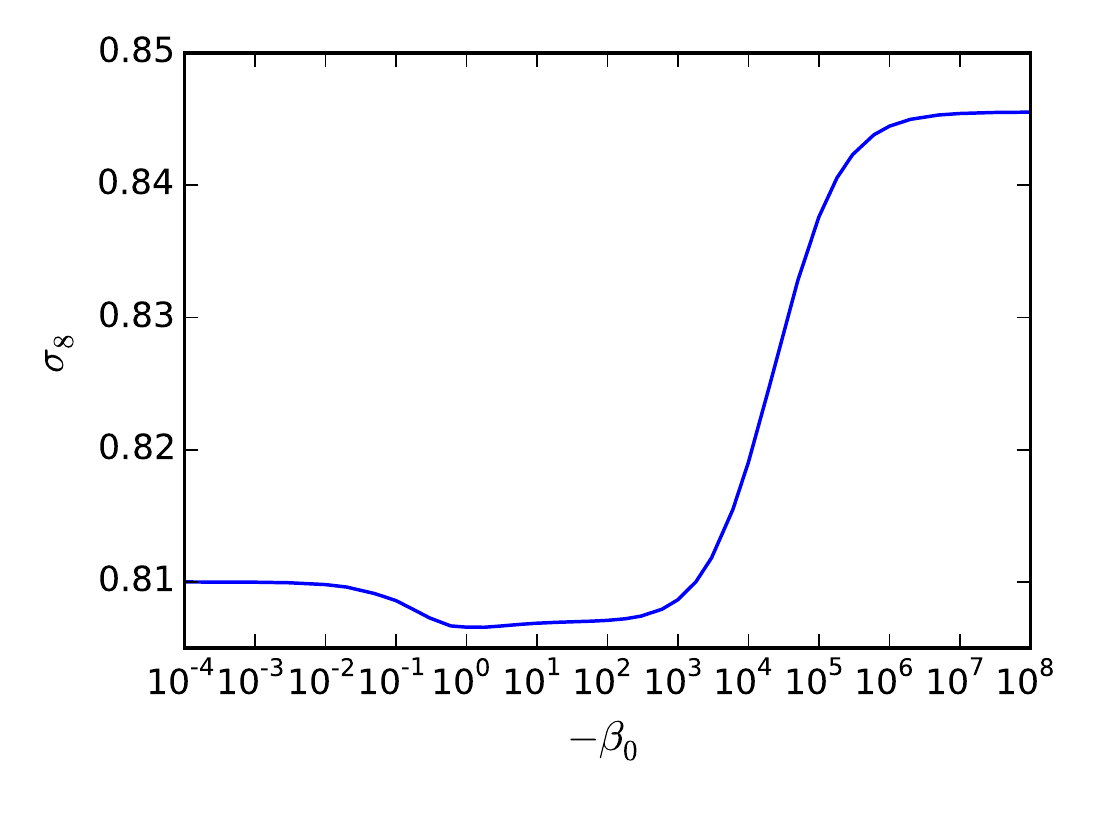}
    \caption{The dependence of $\sigma_8$ on $\beta_0$ for a quadratic coupling function and an exponential potential, as in \cref{fig:0pk}.}
    \label{fig:0sigma8}
\end{figure}

The structure growth suppression is illustrated for the $n=2$ case by \cref{fig:0pk,fig:0sigma8,}, which show the linear matter power spectrum and $\sigma_8$ for the quadratic coupling function and an exponential potential. Following Ref.~\cite{Pourtsidou:2016ico} here we fix the slope of the potential to be $\lambda = 1.22$, which is within the range of values providing a good fit to cosmological data~\cite{Pourtsidou:2016ico}. In \cref{sec:back,sec:perts} we investigate the effect of varying $\lambda$.
We have set the sound horizon at recombination, which is tightly constrained by CMB measurements~\cite{Planck2018VI}, to $\theta_\text{s} = 0.0104$. 
Except where stated otherwise, we keep $\lambda$ and $\theta_\text{s}$ fixed throughout. 
\Cref{fig:0pk} is the analogue of the right-hand panel of Fig.~2 in Ref.~\cite{Pourtsidou:2016ico}. The slightly different value of $P(k)$ in the large $k$ limit is due to different input parameters used in Ref.~\cite{Pourtsidou:2016ico}.
For moderate values of $\beta_0$ there is a slight reduction in $\sigma_8$ relative to uncoupled quintessence (given by the limit of small $|\beta_0|$). For large values of $|\beta_0|$ we see enhancement of $\sigma_8$ relative to uncoupled quintessence. Qualitatively, the suppression arises because the Type~3 coupling gives the CDM fluid a non-zero velocity divergence $\theta_\text{c}$, given by \cref{eq:theta0}. This results in a suppression of the CDM density contrast. We find numerically that the two terms on the right hand side of \cref{eq:delta0} are always of opposite signs, and the second term is larger in magnitude, which means that the larger $|\theta_\text{c}|$, the smaller $|\delta_\text{c}|$ is.


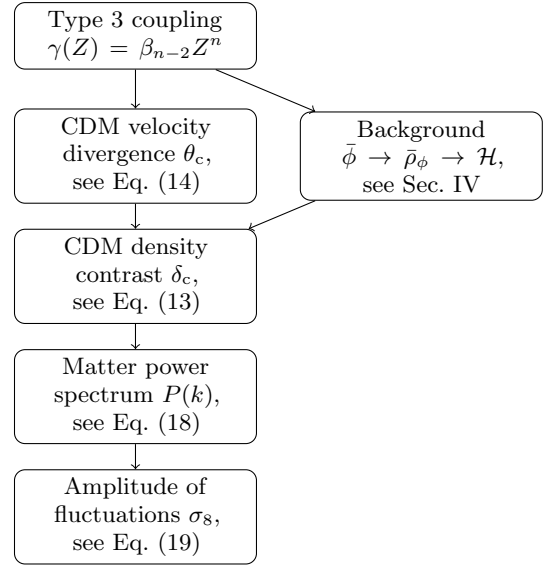
\begin{figure}
\centering
\begin{tikzpicture}[node distance=1.6cm]
    \node (coupling) [chainstep, text width=3cm] {Type 3 coupling $\gamma(Z) = \beta_{n-2}Z^n$};
    \node (theta) [chainstep, below of=coupling, text width=3cm] {CDM velocity divergence $\theta_\text{c}$, \\see \cref{eq:theta0}};
    \node (delta) [chainstep, below of=theta, text width=3cm] {CDM density contrast $\delta_\text{c}$, \\see \cref{eq:delta0}};
    \node (P) [chainstep, below of=delta, text width=3cm] {Matter power spectrum $P(k)$, \\see \cref{eq:P}};
    \node (sigma) [chainstep, below of=P, text width=3cm] {Amplitude of fluctuations $\sigma_8$, \\see \cref{eq:sigma}};
    \node (back) [chainstep, right of=theta, xshift=2.2cm, text width=3cm] {Background $\bar{\phi} \rightarrow \bar{\rho}_\phi \rightarrow \mathcal{H}$, \\see \cref{sec:back}};
    
    \draw[->] (coupling) -- (theta);
    \draw[->] (theta) -- (delta);
    \draw[->] (delta) -- (P);
    \draw[->] (P) -- (sigma);
    \draw[->] (coupling) --(back);
    \draw[->] (back) -- (delta);
\end{tikzpicture}
\caption{A schematic illustration of the steps by which the Type~3 coupling affects the amplitude of fluctuations $\sigma_8$.}
\label{fig:flowchart}
\end{figure}

The steps by which the Type~3 coupling $\gamma(Z)$ impacts the parameter $\sigma_8$ are shown schematically in \cref{fig:flowchart}. In \cref{sec:back} we discuss how the Type~3 coupling affects the background cosmological evolution and in \cref{sec:perts} we demonstrate its effect on the perturbations, in particular how the CDM density contrast $\theta_\text{c}$ depends on the coupling.

\section{Effect of Type~3 Coupling on the Background Evolution}
\label{sec:back}

For a Type~3 coupled quintessence model with power law coupling $\gamma(Z) = \beta_{n-2}Z^n$ and an exponential potential $V(\phi) = A\e{-\lambda\phi/M_\text{P}}$ the scalar field evolution equation, Eq.~(\ref{eq:sfeback}), becomes
\begin{multline}
\label{eq:SFEn}
    \left[1-n(n-1)\beta_{n-2}\left(-\frac{\dot{\bar{\phi}}}{a}\right)^{n-2}\right]\ddot{\bar{\phi}} + 2\mathcal{H}\dot{\bar{\phi}} 
    \\ + n(4-n)a\mathcal{H}\beta_{n-2} \left(-\frac{\dot{\bar{\phi}}}{a}\right)^{n-1} - a^2 \frac{\lambda A}{M_\text{P}} \e{-\lambda\bar{\phi}/M_\text{P}} = 0\,,
\end{multline}
where we have used $\bar{Z} = -\dot{\bar{\phi}}/a$. To understand the behaviour of the scalar field it is instructive to consider certain limiting cases.

In the interest of readability, in the following we use the general quantity $\gamma_{,ZZ}$ in place of its specific form for the power law coupling, $n(n-1)\beta_{n-2}(-\dot{\bar{\phi}}/a)^{n-2}$. 
First we consider the case in which $1 \gg |\gamma_{,ZZ}|$. This can result from either $|\beta_{n-2}|$ or $|\dot{\bar{\phi}}|$ being very small. In this limit, the second term in the square bracket of \cref{eq:SFEn} becomes negligible, as does the third term of the equation and hence
\begin{equation}
\label{eq:SFEnearly}
    \ddot{\bar{\phi}} + 2\mathcal{H}\dot{\bar{\phi}} - a^2 \frac{\lambda A}{M_\text{P}} \e{-\lambda\bar{\phi}/M_\text{P}} = 0\,,
\end{equation}
which is simply the scalar field equation for uncoupled quintessence. In this case the scalar field will roll down the potential with $\bar{\phi}$ and $\dot{\bar{\phi}}$ increasing with time.

In the opposite case, where $1 \ll |\gamma_{,ZZ}|$, \cref{eq:SFEn} becomes
\begin{multline}
\label{eq:SFEnlate}
     -n(n-1)\beta_{n-2}\left(-\frac{\dot{\bar{\phi}}}{a}\right)^{n-2} \ddot{\bar{\phi}} 
     \\+ n(4-n)a\mathcal{H}\beta_{n-2} \left(-\frac{\dot{\bar{\phi}}}{a}\right)^{n-1} - a^2 \frac{\lambda A}{M_\text{P}} \e{-\lambda\bar{\phi}/M_\text{P}} = 0\,.
\end{multline}
Note that for $n=4$ the second term in \cref{eq:SFEnlate} is equal to zero and therefore one would not neglect the $2\mathcal{H}\dot{\bar{\phi}}$ term in \cref{eq:SFEn}. For our present purposes, however, this distinction is not vital. What is important to note is that, since we are considering the regime where $|\gamma_{,ZZ}| \gg 1$, \cref{eq:SFEnlate} predicts a slower evolution of $\dot{\bar{\phi}}$ and hence $\bar{\phi}$ due to the large factor multiplying $\ddot{\bar{\phi}}$.

\begin{figure*}
    \centering
    \includegraphics[width=\columnwidth]{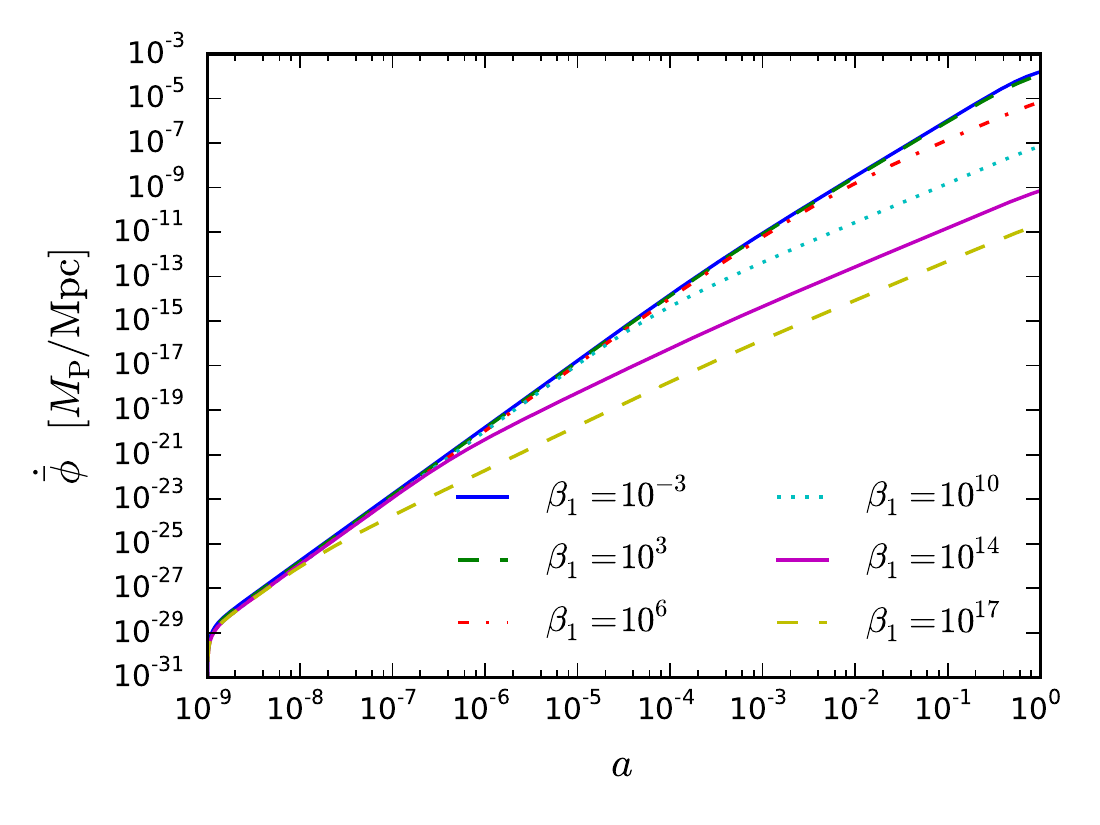}
    \includegraphics[width=\columnwidth]{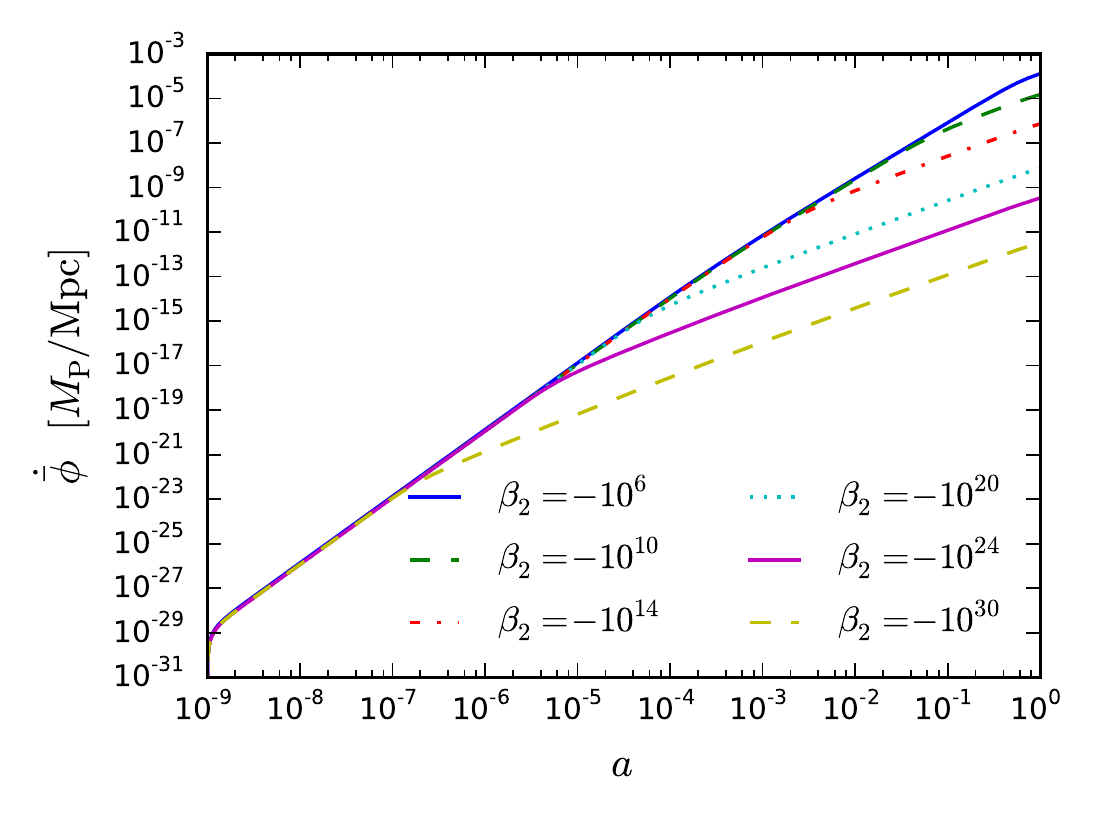}
    \caption{The evolution of the conformal time derivative of the scalar field with the scale factor $a$ for a Type~3 coupling of the form $\gamma(Z) = \beta_{n-2} Z^n$, for different values of the coupling parameter $|\beta_{n-2}|$. The left panel shows the $n=3$ case and the right panel shows $n=4$.
    The units of $\beta_{n-2}$ are $(\mathrm{Mpc}/M_\text{P})^{n-2}$.}
    \label{fig:phip}
\end{figure*}

Provided $n\neq 2$, we can now see that a Type~3 coupled quintessence model will transition from the first limit, when $\dot{\bar{\phi}}$ is very small, to the second limit, since $\dot{\bar{\phi}}$ grows with time. Larger $|\beta_{n-2}|$ will result in an earlier transition from the uncoupled quintessence regime of \cref{eq:SFEnearly} to the `slowed' regime of \cref{eq:SFEnlate}. This is demonstrated by \cref{fig:phip}, which shows the evolution of $\dot{\bar{\phi}}$ for $n=3$ and $n=4$.

To understand the way in which $\dot{\bar{\phi}}$ scales with $\beta_{n-2}$ it is instructive to consider the special case in which $n=2$. In this case, the scalar field equation \cref{eq:SFEn} becomes:
\begin{equation}
    \label{eq:sfe1}
    (1-2\beta_0)(\ddot{\bar{\phi}} + 2\mathcal{H}\dot{\bar{\phi}}) - a^2 \frac{\lambda A}{M_\text{P}} \e{-\lambda\bar{\phi}/M_\text{P}} = 0\,.
\end{equation}
The factor multiplying the kinetic term is now independent of time, implying that the transition described above for general $n$ is not present for $n=2$. Instead, one can see that for small $|\beta_0|$ the uncoupled quintessence case is recovered for all time, and for large $|\beta_0|$ the scalar field evolution is slowed; while for all $\beta_0$, $\dot{\bar{\phi}}$ scales as $1/(1-2\beta_0)$, which we have confirmed  numerically.

We can obtain the scaling behaviour for general $n$, illustrated by \cref{fig:phip}, in a schematic way as follows. 
In analogy to the $n=2$ case in which $\dot{\bar{\phi}}$ scales with $1/(1-2\beta_0)$, let us suppose that $\dot{\bar{\phi}}$ for general $n$ will scale as the inverse of the term in square brackets of \cref{eq:SFEn}:
\begin{equation}
\label{eq:nscale}
    \dot{\bar{\phi}} \sim \frac{1}{1-n(n-1)\beta_{n-2}(-\dot{\bar{\phi}}/a)^{n-2}}\,.
\end{equation}
This relation is of little use in its present form because it contains $\dot{\bar{\phi}}$ on both sides. However, as above, we can consider the two limits: $1 \gg |\gamma_{,ZZ}|$ and $1 \ll |\gamma_{,ZZ}|$. In the first case there is no scaling with $\beta_{n-2}$ as the uncoupled quintessence case is recovered. In the second case, however, \cref{eq:nscale} becomes:
\begin{equation}
\label{eq:nscalelate}
    \dot{\bar{\phi}} \sim \frac{1}{-n(n-1)\beta_{n-2}(-\dot{\bar{\phi}}/a)^{n-2}}\,,
\end{equation}
and we can now rearrange to obtain
\begin{equation}
\label{eq:nscalelate2}
    \dot{\bar{\phi}} \sim |\beta_{n-2}|^{-\frac{1}{n-1}}\,,
\end{equation}
which agrees with the late time, large $|\beta_{n-2}|$ regime in \cref{fig:phip}.

\subsection{Impact on expansion rate}

In order to understand how the expansion rate depends on the Type~3 coupling it is necessary to consider the evolution of the scalar field energy density $\bar{\rho}_\phi$. The energy density and pressure are given by \cref{eq:rhophi,eq:pphi}. For a power-law coupling and exponential potential:
\begin{equation}
\label{eq:rhophi2}
    \bar{\rho}_\phi = \frac{1}{2} \left(\frac{\dot{\bar{\phi}}}{a}\right)^2 - (n-1) \beta_{n-2} \left(-\frac{\dot{\bar{\phi}}}{a}\right)^n + A\e{-\lambda\bar{\phi}/M_\text{P}}\,,
\end{equation}
and 
\begin{equation}
    \label{eq:pphi2}
    \bar{p}_\phi = \frac{1}{2}\left(\frac{\dot{\bar{\phi}}}{a}\right)^2 - \beta_{n-2}\left(-\frac{\dot{\bar{\phi}}}{a}\right)^n - A\e{-\lambda\bar{\phi}/M_\text{P}}\,.
\end{equation}
The energy density of the scalar field obeys the usual conservation equation:
\begin{align}
    &\dot{\bar{\rho}}_\phi + 3\mathcal{H}(\bar{\rho}_\phi + \bar{p}_\phi) = 0 \,,
    \\\Rightarrow \quad &\dot{\bar{\rho}}_\phi = -3\mathcal{H} \left[\left(\frac{\dot{\bar{\phi}}}{a}\right)^2 - n\beta_{n-2}\left(-\frac{\dot{\bar{\phi}}}{a}\right)^n\right]\,.
    \label{eq:T3rhophicons}
\end{align}
Once again it is instructive to consider this expression in the small and large $\beta_{n-2}$ limits separately. In the limit where $1 \gg |\gamma_{,ZZ}|$, the second term in the square bracket of \cref{eq:T3rhophicons} is negligible and the uncoupled quintessence case is recovered. In the limit where $1 \ll |\gamma_{,ZZ}|$, however, the first term in the square bracket can be neglected and one obtains
\begin{equation}
    \dot{\bar{\rho}}_\phi = 3n\mathcal{H} \beta_{n-2}\left(-\frac{\dot{\bar{\phi}}}{a}\right)^n\,.
\end{equation}
We have already established that in this limit, $\dot{\bar{\phi}}$ scales according to \cref{eq:nscalelate2}, so we can infer $\dot{\bar{\rho}}_\phi$ scales with $\beta_{n-2}$ as
\begin{equation}
    \dot{\bar{\rho}}_\phi \sim |\beta_{n-2}|^{-\frac{1}{n-1}}\,.
\end{equation}
Finally, since $\beta_{n-2}(-\dot{\bar{\phi}}/a)^n$ is always negative for the choices of $\beta_{n-2}$ we consider, we can conclude that $\bar{\rho}_\phi$ falls with time, and does so more slowly the larger $|\beta_{n-2}|$ is. Thus for very large values of $|\beta_{n-2}|$ the scalar field behaves similarly at the background level to a cosmological constant.

\begin{figure}
    \centering
    \includegraphics[width=\columnwidth]{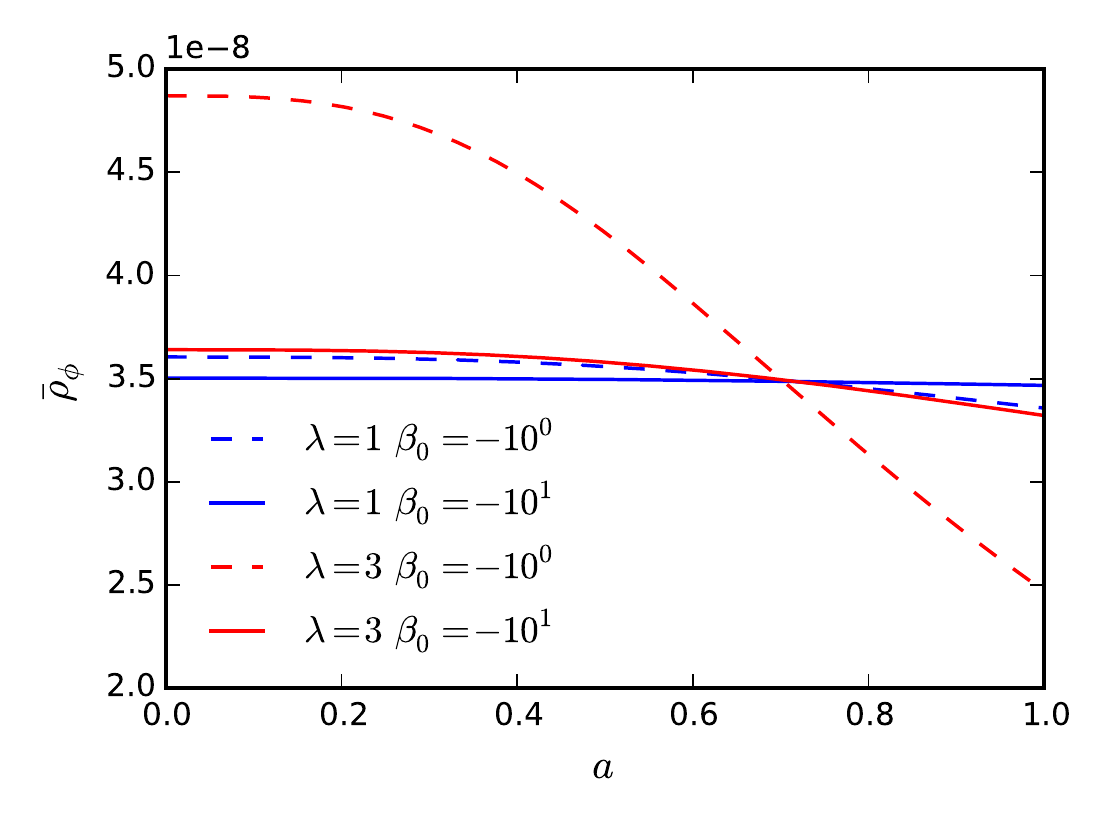}
    \caption{The evolution of the background energy density of the scalar field $\bar{\rho}_\phi$ as a function of the scale factor,$a$, for two values of the coupling parameter $\beta_0$ and the potential parameter $\lambda$. 
    }
    \label{fig:rhophi}
\end{figure}

As in the case of uncoupled quintessence, a steeper scalar field potential also results in a faster evolution of $\bar{\phi}$ and hence a drop in $\bar{\rho}_\phi$. In terms of the background evolution we therefore find that the potential parameter $\lambda$ and the Type~3 coupling parameter $\beta_{n-2}$ act in opposition to each other, with an increase in the former tending to speed up the scalar field evolution and the latter tending to slow it. Both of these effects are illustrated by \cref{fig:rhophi} for a Type~3 coupling with $n=2$.

\begin{figure}
    \centering
    \includegraphics[width=\columnwidth]{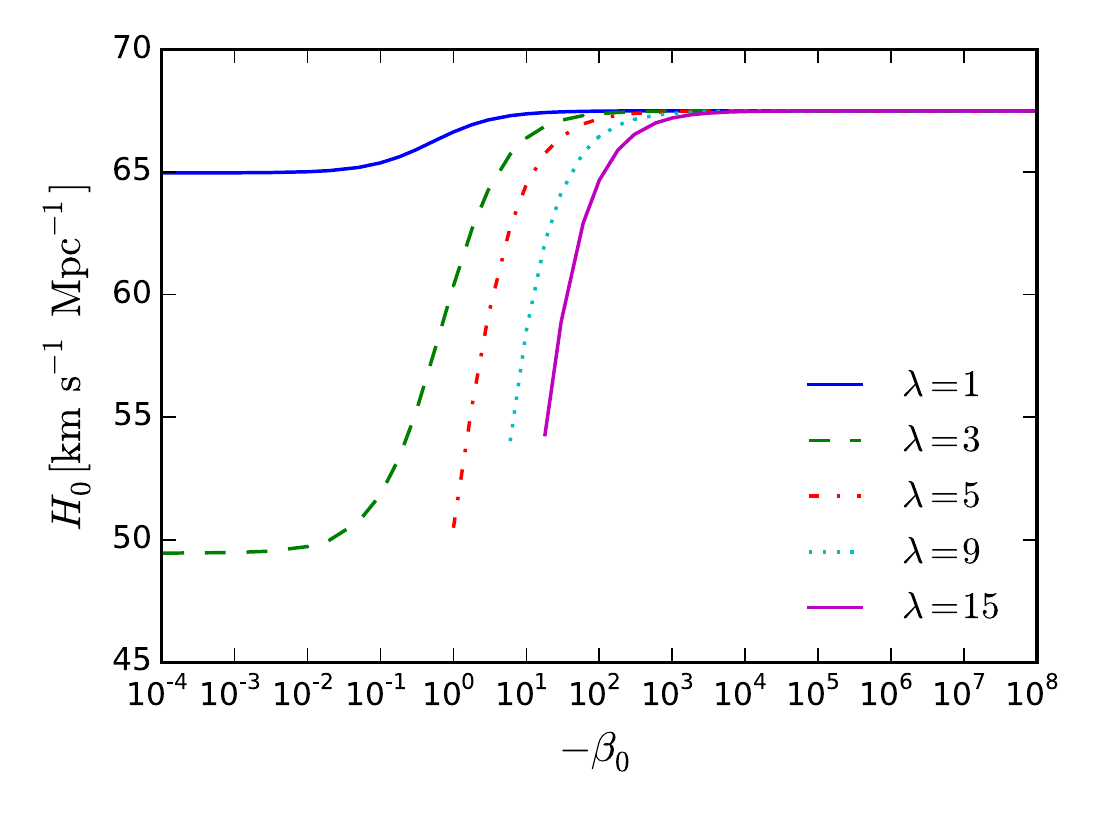}
    \caption{The present-day value of the Hubble parameter, $H_0$, as a function of the coupling parameter, $|\beta_0|$, for a range of potential parameters, $\lambda$, for a quadratic coupling function $\gamma(Z) = \beta_0 Z^2$ and an exponential potential $V(\phi) = A \e{-\lambda\phi/M_\text{P}}$.}
    \label{fig:h}
\end{figure}

The expansion rate can be calculated using the Friedmann equation. At early times, the contribution of the scalar field is negligible compared to those of matter and radiation but at late times it is the dominant species. A small value of $|\beta_{n-2}|$, allowing $\bar{\rho}_\phi$ to fall, will give rise to a smaller present-day expansion rate than a large value of $|\beta_{n-2}|$, which slows the evolution of $\bar{\rho}_\phi$ and gives an expansion rate close to that expected from a cosmological constant. \Cref{fig:h} illustrates the impact of the slope of the potential and the Type~3 coupling parameter on the present-day expansion rate for a Type~3 coupling with $n=2$. It can be seen that steep potentials give rise to unrealistically low values of $H_0$ unless the Type~3 coupling is sufficiently strong.

\section{Evolution of linear perturbations}
\label{sec:perts}

\subsection{Dependence on coupling parameter}

Type~3 models affect the cosmological perturbations through the modified equation for the CDM velocity divergence, \cref{eq:theta0}. In the case of a power-law coupling, \cref{eq:theta0} can be written:
\begin{equation}
\label{eq:thetacn}
    \dot{\theta}_\text{c} + \mathcal{H}\theta_\text{c} = \frac{n\beta_{n-2} [a^{4-n} (-\dot{\bar{\phi}})^{n-1} \delta\phi] \dot{}} {a^4[\bar{\rho}_\text{c}-n\beta_{n-2}(-\dot{\bar{\phi}}/a)^n]}\,.
\end{equation}
It turns out that the second term in the denominator is always significantly smaller than the first. Thus to understand the behaviour of $\theta_\text{c}$ it suffices to consider the numerator. Let us separately consider how $\beta_{n-2}(-\dot{\bar{\phi}})^{n-1}$ and $\delta\phi$ depend on $\beta_{n-2}$. 

From \cref{eq:nscalelate2} we can see that, for sufficiently large $|\beta_{n-2}|$, the factor $\beta_{n-2}(-\dot{\bar{\phi}})^{n-1}$ is approximately constant and any $\beta$-dependence of $\theta_\text{c}$ must come from $\delta\phi$. In the limit of small $|\beta_{n-2}|$, however, we have already seen that $\dot{\bar{\phi}}$ is approximately independent of $\beta_{n-2}$ so the factor $\beta_{n-2}(-\dot{\bar{\phi}})^{n-1}$ rises linearly with $\beta_{n-2}$.

\begin{figure}
    \centering
    \includegraphics[width=\columnwidth]{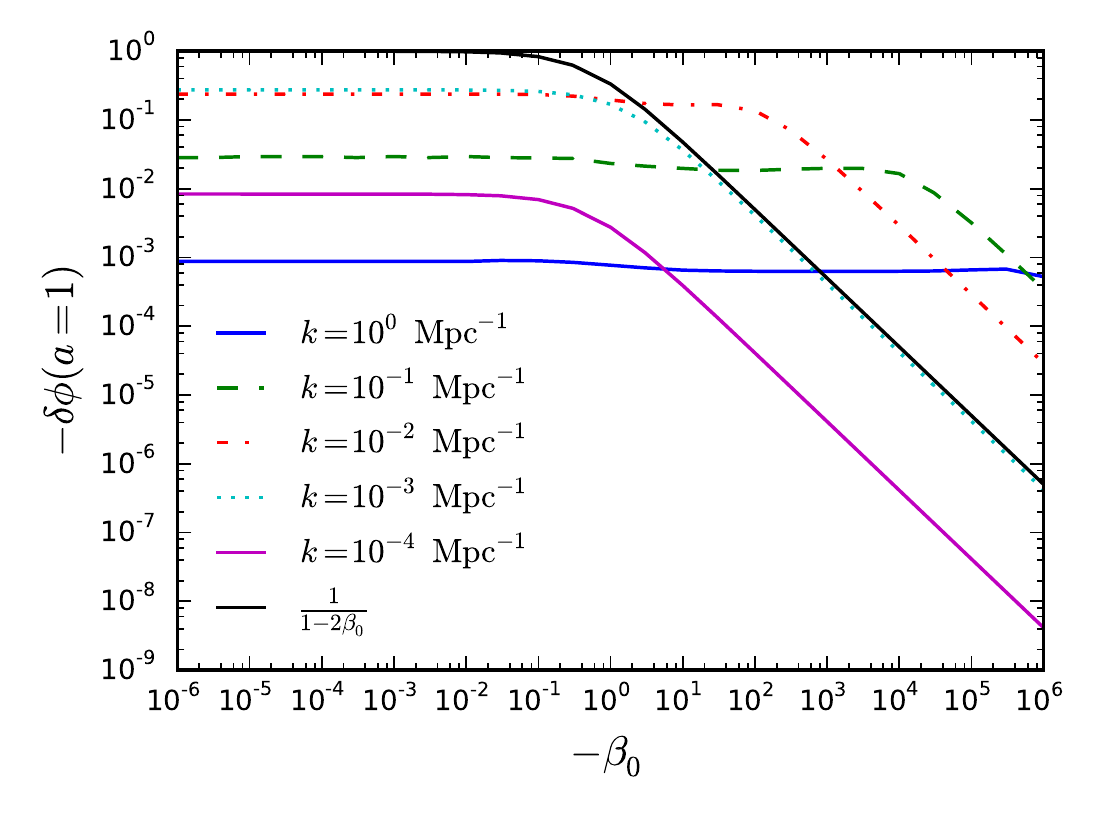}
    \includegraphics[width=\columnwidth]{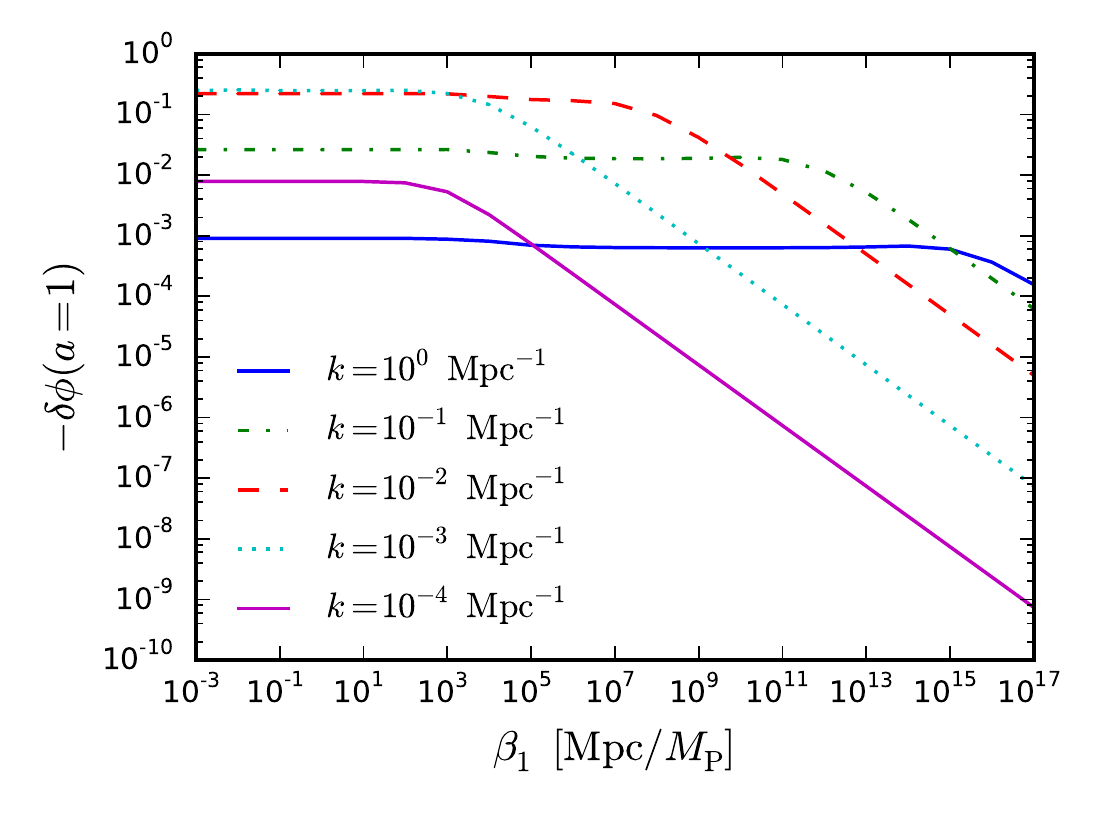}
    \caption{The present-day value of the scalar field perturbation $\delta\phi$ for a Type~3 coupling $\gamma(Z) = \beta_{n-2}Z^n$ as a function of $\beta_{n-2}$ for different scales $k$. The top panel shows the $n=2$ case and the bottom panel shows $n=3$. 
    }
    \label{fig:dphi}
\end{figure}

As illustrated by \cref{fig:dphi}, $\delta\phi$ is constant with $\beta_{n-2}$ for small $|\beta_{n-2}|$, and drops as 
\begin{equation}
    \delta\phi \sim |\beta_{n-2}|^{-\frac{1}{n-1}}\,,
\end{equation}
for large $|\beta_{n-2}|$, with the transition from the approximately constant regime to the $|\beta_{n-2}|^{-1/(n-1)}$ regime occurring at larger $|\beta_{n-2}|$ on smaller scales. For the special case of $n=2$, we can make a more precise statement and say that $\delta\phi$ scales as $1/(1-2\beta_0)$ on large scales. The black line in the top panel of \cref{fig:dphi} illustrates this scaling, closely matching the form of the magenta and cyan lines which correspond to large scales, while the blue line, corresponding to small scales, is constant for a wide range of $\beta_0$. 

\begin{figure}
    \centering
    \includegraphics[width=\columnwidth]{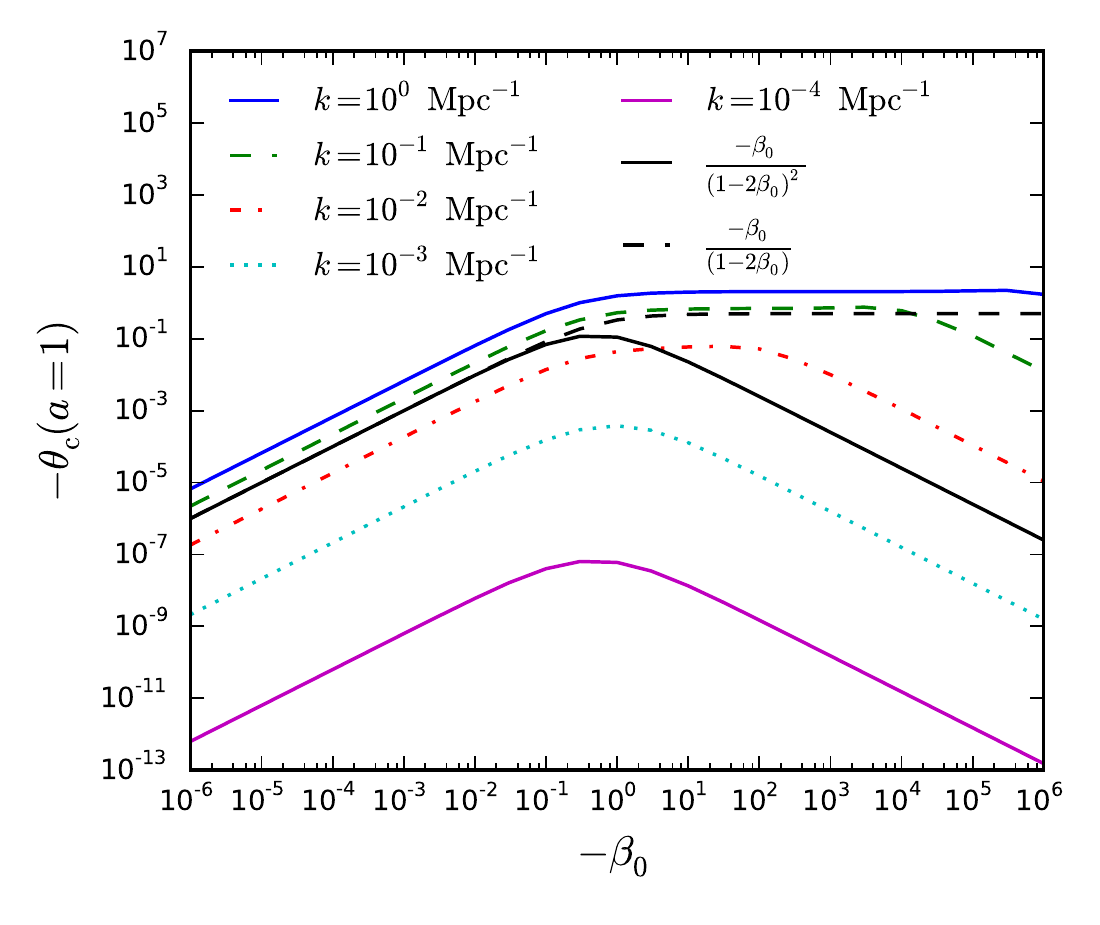}
    \includegraphics[width=\columnwidth]{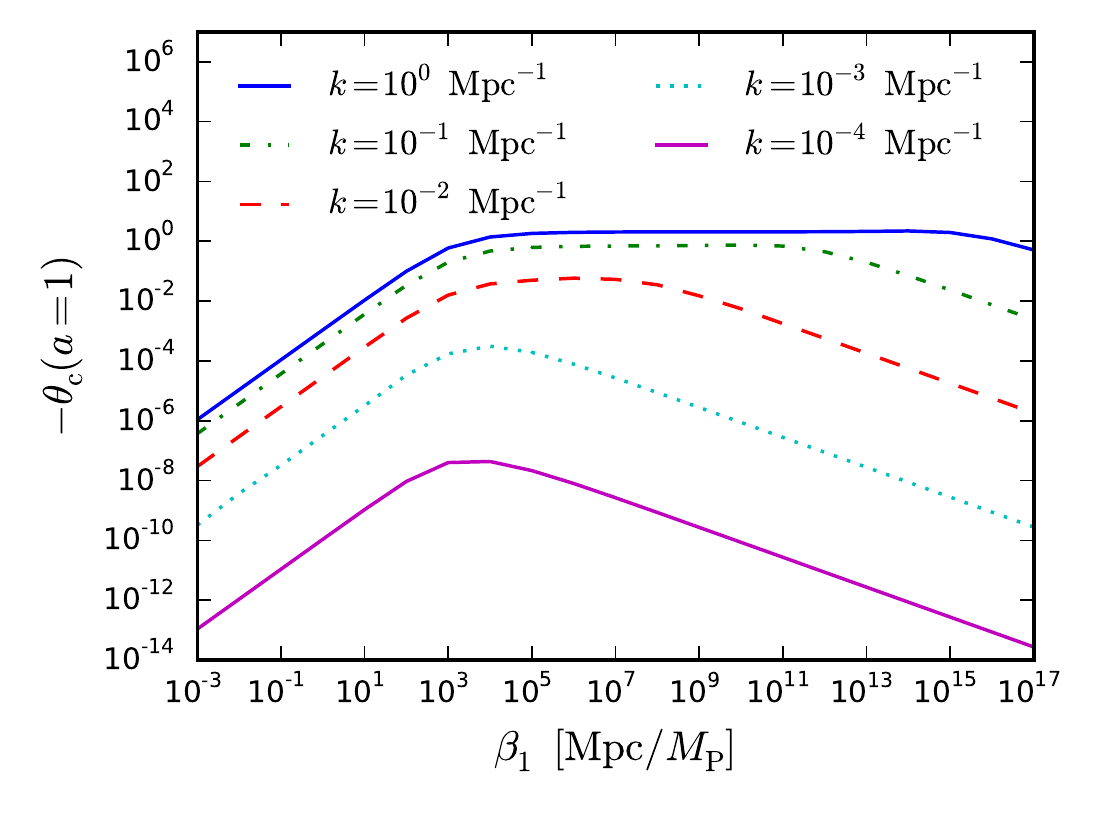}
    \caption{The present-day CDM velocity divergence $\theta_\text{c}$ for a Type~3 coupling $\gamma(Z) = \beta_{n-2}Z^n$ as a function of $\beta_{n-2}$ for several scales $k$. The top panel shows the $n=2$ case and the bottom panel shows $n=3$.}
    \label{fig:thetac}
\end{figure}

The above scaling arguments for the factors in \cref{eq:thetacn} allow us to understand how $\theta_\text{c}$ depends on $\beta_{n-2}$. For small $|\beta_{n-2}|$, we expect $|\theta_\text{c}|$ to rise linearly with $|\beta_{n-2}|$, while for large $|\beta_{n-2}|$ we expect it to fall as $|\beta_{n-2}|^{-1/(n-1)}$. This behaviour is illustrated in \cref{fig:thetac}. The broad peak of $|\theta_\text{c}|$ on small scales results from the fact that $\delta\phi$ is approximately constant for a wide range of $\beta_{n-2}$ on small scales. Once again, we can be more precise in the special case in which $n=2$. Inserting the scalings for $\dot{\bar{\phi}}$ and $\delta\phi$ into \cref{eq:thetacn}, we find that $\theta_\text{c}$ scales as $\beta_0/(1-2\beta_0)^2$ on large scales (black solid line in the top panel of \cref{fig:thetac}) and as $\beta_0/(1-2\beta_0)$ on small scales (black dashed line in the top panel of \cref{fig:thetac}).

\subsection{Dependence on potential}

\begin{figure}
    \centering
    \includegraphics[width=\columnwidth]{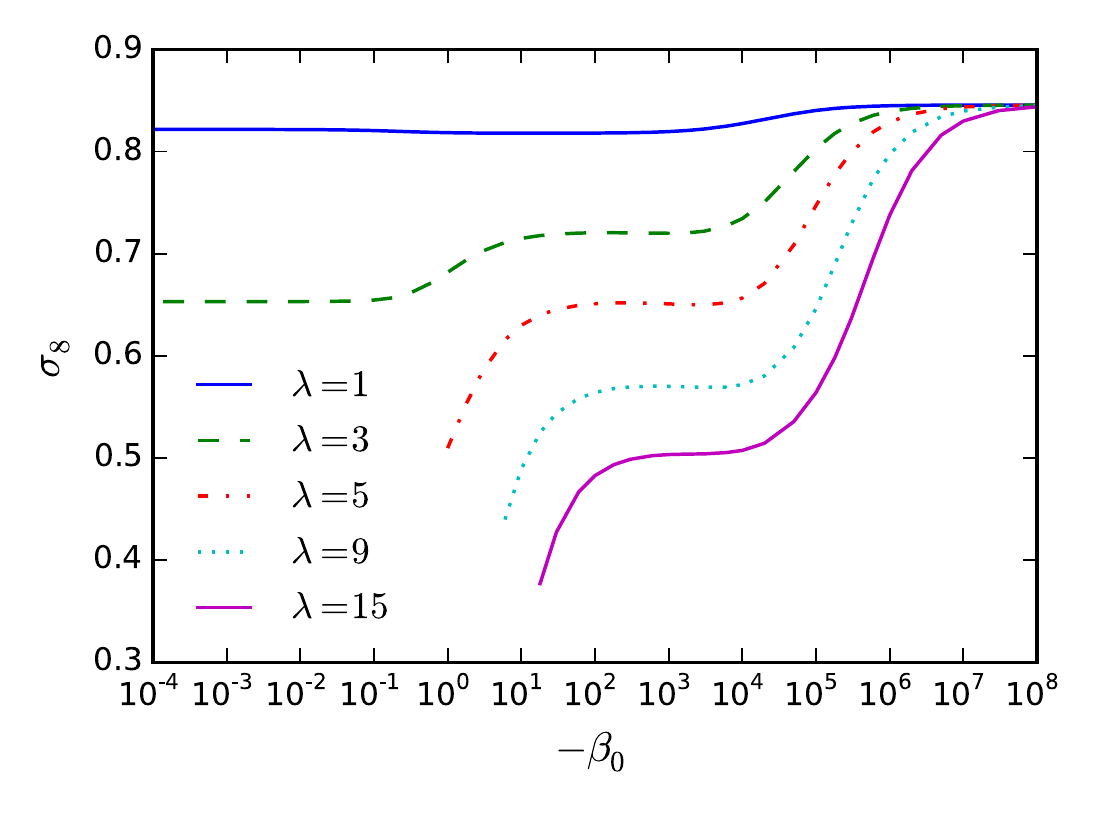}
    \caption{The amplitude of matter fluctuations $\sigma_8$ as a function of the coupling parameter $|\beta_0|$ for a range of potential parameters $\lambda$ for a quadratic coupling function $\gamma(Z) = \beta_0 Z^2$ and an exponential potential $V(\phi) = A \e{-\lambda\phi/M_\text{P}}$.}
    \label{fig:sigma8}
\end{figure}

The slope of the potential, $\lambda$, also has an impact on the growth of matter perturbations. \Cref{fig:sigma8} demonstrates for a coupling with $n=2$ that a steeper potential can give rise to a very large reduction in $\sigma_8$, certainly large enough to resolve the discrepancy between early- and late-universe observations. It should be noted that increasing the slope $\lambda$ of the potential has the effect of reducing the expansion rate, which is to be avoided since this exacerbates the Hubble tension. (See Ref.~\cite{Verde:2019ivm} for a recent discussion.) The lines in \cref{fig:sigma8} stop once  $H_0< 30 \, \mathrm{km}\mathrm{s}^{-1}\mathrm{Mpc}^{-1}$ but even much smaller reductions in $H_0$ are problematic. However, from \cref{fig:h,fig:sigma8} one can see that there are choices for $\lambda$ and $\beta_0$ that give rise to significant reduction in $\sigma_8$ without having a noticeable impact on $H_0$, for example $\lambda=3$, $\beta_0 = -10^2$.

\begin{figure}
    \centering
    \includegraphics[width=\columnwidth]{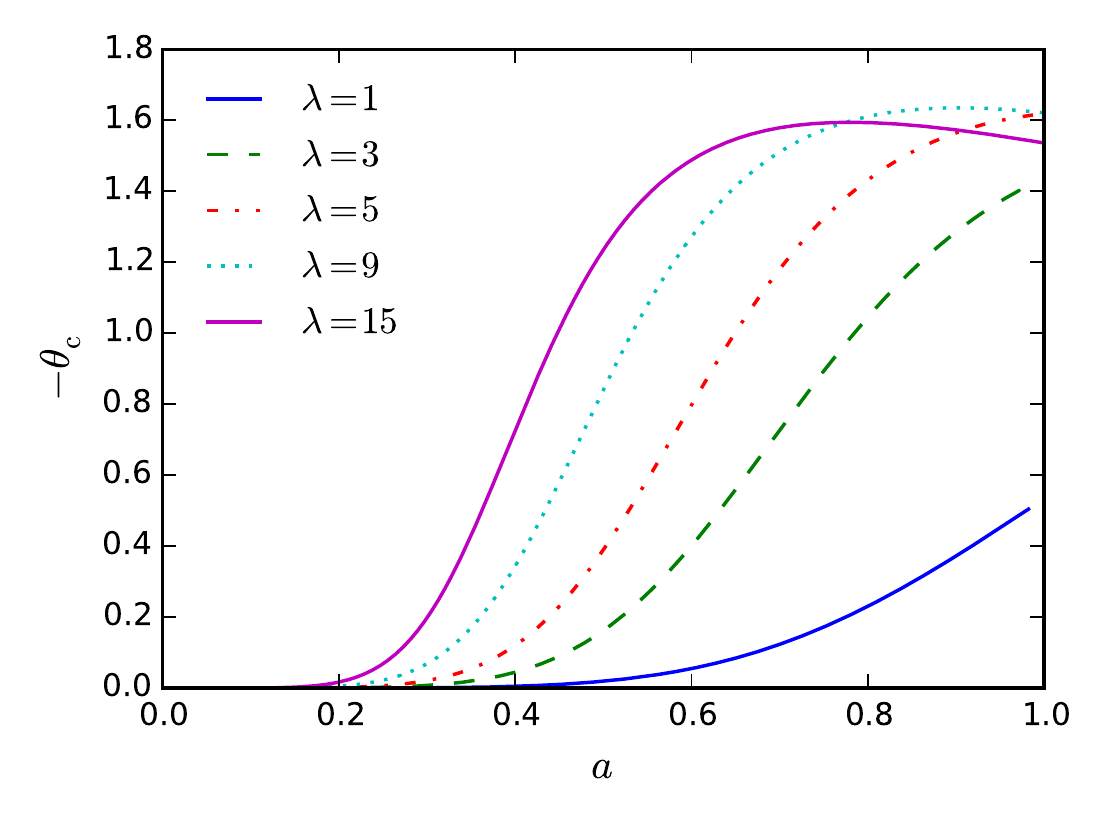}
    \caption{The evolution of the CDM velocity divergence $\theta_\text{c}$ as a function of the scale factor $a$ for a range of different potential parameters $\lambda$ with a coupling parameter $\beta_0 = -10^2$, at a scale $k = 0.1 \,\mathrm{Mpc}^{-1}$. 
    The sound horizon at recombination is held fixed at $\theta_\text{s} = 0.0104$.}
    \label{fig:thetac_lambda}
\end{figure}

To understand how structure growth depends on $\lambda$ one needs to consider the CDM velocity divergence. \Cref{fig:thetac_lambda} shows, again for the $n=2$ case, how the evolution of $\theta_\text{c}$ is affected by the potential parameter $\lambda$: larger $\lambda$, corresponding to a steeper potential, results in $|\theta_\text{c}|$ rising more rapidly. Larger $\theta_\text{c}$ at a given time reduces the time derivative of the CDM density contrast $\delta_\text{c}$ (see \cref{eq:delta0}), resulting in a smaller $|\delta_\text{c}|$ at the present epoch and hence a reduction of $\sigma_8$ for large $\lambda$ as seen in \cref{fig:sigma8}.
The $\lambda$-dependence of $\theta_\text{c}$ can be seen in the $\theta_\text{c}$ equation (\cref{eq:theta0}). Substituting for
$\ddot{\bar{\phi}}$ using \cref{eq:sfe1}, \cref{eq:theta0} becomes
\begin{equation}
    \label{eq:thetapot}
    \dot{\theta}_\text{c} = -\mathcal{H}\theta_\text{c} + \frac{\frac{2\beta_0}{1-2\beta_0} a^2 V_{,\phi} \delta\phi - 2\beta_0\dot{\bar{\phi}}\dot{\delta\phi}} {(\bar{\rho}_\text{c}a^2 - 2\beta_0\dot{\bar{\phi}}^2)}\,,
\end{equation}
which, for an exponential potential $V(\phi) = A\e{-\lambda\phi/M_\text{P}}$, yields
\begin{equation}
    \label{eq:thetapot2}
    \dot{\theta}_\text{c} = -\mathcal{H}\theta_\text{c} + \frac{-\frac{2\beta_0}{1-2\beta_0} a^2 A\lambda\e{-\lambda\phi/M_\text{P}}\delta\phi - 2\beta_0\dot{\bar{\phi}}\dot{\delta\phi}} {(\bar{\rho}_\text{c}a^2 - 2\beta_0\dot{\bar{\phi}}^2)}\,.
\end{equation}
Both of the terms in the numerator become larger in magnitude when $\lambda$ is large. In the first term this is obvious; in the second it is a consequence of the $V_{,\phi\phi}$ term in \cref{eq:sfeperturb}. Hence, a large slope $\lambda$ results in a large (negative) $\theta_\text{c}$ leading to a reduction in $\delta_\text{c}$ and a suppression of structure growth.

\subsection{Metric perturbation}
As can be seen in \cref{eq:delta0}, the CDM density contrast depends not only on the CDM velocity divergence but also on the time derivative of the metric perturbation $h$. This latter quantity has a weak indirect dependence on the Type~3 coupling through its dependence on the background expansion rate. In general, a larger expansion rate at a given time results in a smaller value of $\dot{h}$, which in turn reduces $|\delta_\text{c}|$ and suppresses structure growth. For all of the cases we have considered, this effect is much smaller than the effect due to $\theta_\text{c}$.


\section{Conclusions}
\label{sec:disc}

Unlike most coupled dark energy models that have been studied in the literature, Type~3 models, as classified at the Lagrangian level in Ref.~\cite{Pourtsidou:2013nha}, consist of a coupling between the momentum of the dark matter and the gradient of the dark energy scalar field. It was demonstrated in Ref.~\cite{Pourtsidou:2016ico} using MCMC methods that such models can ease the tension between early- and late-universe measurements of the degree of structure formation in the universe. 

In this work we have presented an explanation, using both analytical and numerical methods, of why Type~3 models suppress the growth of structure. We considered a fairly general power-law coupling function, finding that it gives rise to similar structure suppression behaviour to the quadratic case previously studied. 

We explored in detail the behaviour of the background cosmological evolution of Type~3 coupled quintessence models, demonstrating how the scalar field evolution depends on the coupling and the scalar field potential, and how the expansion rate is affected. In particular, we find that, as with uncoupled quintessence, a steeper slope of the scalar field potential gives rise to faster evolution of the scalar field and thus a reduced present-day expansion rate $H_0$. On the other hand, increasing the strength of the coupling via the parameter $\beta_{n-2}$ slows down the scalar field evolution and gives rise to a present-day expansion rate similar to that predicted by \lcdm{}. 
In the models we have studied there is no mechanism for increasing $H_0$ beyond the value predicted by \lcdm{} and hence resolving the existing Hubble constant tension.

This understanding of the background evolution was then applied to the perturbed equations of motion. A Type~3 coupling between dark energy and CDM gives rise to a non-zero CDM velocity divergence, $\theta_\text{c}$, which suppresses structure growth via its role in the evolution of the density contrast of CDM, $\delta_\text{c}$ (see \cref{eq:delta0}). We found that the value of $|\theta_\text{c}|$ rises and falls with $|\beta_{n-2}|$, with a maximum corresponding to the maximum possible structure suppression. We demonstrated this behaviour using both approximate analytic arguments based on the equations of motion and a numerical analysis using an appropriately modified version of \CLASS{}.

We also demonstrated how the structure suppression depends on the slope $\lambda$ of the scalar field potential. In particular, increasing $\lambda$ gives rise to a stronger suppression of structure growth. As our background analysis demonstrated, this can have the unwanted side effect of reducing the predicted value of $H_0$, thus worsening the Hubble tension. However, for appropriate values, such as $\lambda = 3$ and $\beta_0 = -10^2$, the structure suppression can be achieved without the Hubble constant being reduced. Thus our results indicate an even greater suppression of structure formation is possible than what has previously been realised.

In order to understand the physical origin of the suppression of structure, we have held most cosmological parameters fixed.
To fully explore the interplay between model parameters such as $\beta_{n-2}$ and $\lambda$ and cosmological parameters such as $\sigma_8$ and $H_0$ a multi-parameter MCMC analysis is needed. In Ref.~\cite{Linton:2017ged} such an analysis was carried out using CMB data from Planck for a Type~3 model with a cubic coupling, allowing the potential parameter $\lambda$ to vary between 0 and 2.1. They found the Type~3 model to be consistent with the CMB data but marginally disfavoured when compared to \lcdm{}.

We have not discussed the physical origin of the Type~3 coupling. 
Presenting a more physically motivated model would be a worthwhile avenue for future study. Recently in \cite{Kase:2019veo,Kase:2019mox} the authors have considered the presence of related interactions in the context of Horndeski theories of modified gravity. We note that our analysis has involved the use of large dimensionless numbers for the coupling parameter $\beta_0$. Without reference to a deeper underlying theory it is difficult to say whether such values are reasonable, but the requirement of large dimensionless numbers is somewhat unappealing. This would be a challenge for any future physically motivated Type~3 theory. Another possible focus of future research would be to study Type~3 models in a more model-independent way, using the PPF formalism developed in Ref.~\cite{Skordis:2015yra}. In this approach, there is a certain set of non-zero parameters that define a Type~3 model, which can in principle be constrained by observational surveys. Type~3 interacting dark energy is still a young and little-studied area of research but it has been shown to have interesting consequences for the structure and evolution of the universe.


\section{Acknowledgements}
F.N.C. is supported by a United Kingdom Science and Technology Facilities Council (STFC) studentship.  A.A., E.J.C. and A.M.G. acknowledge  support  from  STFC  grant ST/P000703/1. A.P. is a UK Research and Innovation Future Leaders Fellow, grant MR/S016066/1, and also acknowledges support from the UK Science and Technology Facilities Council through grant ST/S000437/1. A.P. is grateful to the University of Nottingham for hospitality during the initial stages of this work. We are grateful to Thomas Tram for useful discussions. 

\bibliography{T3paper.bib}

\end{document}